\documentclass[final,preprint,5p,times,twocolumn, authoryear]{elsarticle}

\usepackage{amssymb}

\usepackage{amsmath}

\journal{arXiv.org}

\begin{document}

\begin{frontmatter}



\title{Russian Agricultural Industry under Sanction Wars}
\author{Alexandra Lukyanova}
\author{Ayaz Zeynalov\cortext[]{\scriptsize Corresponding author: Ayaz Zeynalov, Prague University of Economics and Business, Faculty of International Relations. Address: , 130 67, Prague, Czechia. T: (+420) 224 095 239; \textit{Email:} ayaz.zeynalov@vse.cz. Alexandra Lukyanova is student at Prague University of Economics and Business, \textit{Email:} luka02@vse.cz. Financial support from the Prague University of Economics and Business (grant IG212021) is gratefully acknowledged.}}

\affiliation{organization={Prague University of Economics and Business},
 addressline={W. Churchill Sq. 1938/4},
 city={Prague},
 postcode={130 67},
 country={Czechia}}

\begin{abstract}
	The motivation for focusing on economic sanctions is the mixed evidence of their effectiveness. We assess the role of sanctions on the Russian international trade flow of agricultural products after 2014. We use a differences-in-differences model of trade flows data for imported and exported agricultural products from 2010 to 2020 in Russia. The main expectation was that the Russian economy would take a hit since it had lost its importers. We assess the economic impact of the Russian food embargo on agricultural commodities, questioning whether it has achieved its objective and resulted in a window of opportunity for the development of the domestic agricultural sector. Our results confirm that the sanctions have significantly impacted foodstuff imports; they have almost halved in the first two years since the sanctions were imposed. However, Russia has embarked on a path to reduce dependence on food imports and managed self-sufficient agricultural production.
\end{abstract}




\end{frontmatter}

\section{Introduction}
\label{sec:intro}

Recently, sanctions have become an increasingly common means of influence in foreign policy, as is growing evidence in the media and academic literature. The motivation for focusing on economic sanctions is the mixed evidence of their effectiveness. Literature has been focused on assessing the impact of sanctions on the economy of the concerned country; however, the effectiveness of sanctions is still open for discussion. The success rate of sanctions was determined based on whether the country or organisation initiating the sanctions achieved its declared goals. \citet{Hufbauer1990} emphasise that the sanctions work best for friendly countries (50\% success - against friendly countries, 33\% - against neutral countries and 19\% - against hostile countries). The authors claim that sanctions work in only 34 of the 100 cases analysed \citep{Hufbauer2009}. At the same time, very rarely have sanctions succeeded in stopping military aggression by hostile countries. According to \citet{Pape1997}, the sanctions do not contribute to key international policy goals. \citet{Hufbauer2020} revisited the impacts of sanctions and explained the different perspectives of sanctions for different countries. 

Russia relied heavily on importing all its fresh and transformed food from the international market, especially from the western world, before 2014. Several waves of sanctions have introduced economic restrictions on Russian financial institutions, senior politicians, and multinational companies after the Crimea crisis. These sanctions have been punctual and imposed to put pressure on Russia to change its stance on major international issues and weaken the Russian economy. Russia was understandably not going to stand by, and, in a tit-for-tat action, has taken a bold action by imposing sanctions on the western world by banning almost all foodstuff products (i.e., meats, dairy, fruits \& vegetables, and seafood) from being imported. 

Since the imposition of the embargo, economists have debated the consequences of Russia's food import ban. In assessing the consequences of the embargo, both positive and negative aspects for the Russian food market have been noted. The main expectation was that the Russian economy would take a hit since it lost its importers. However, this action was not taken by chance; the issue of food security in the country has been a priority since 2007. It is known that the country's heavy dependence on food imports poses a significant threat to the country's food security, and this problem has to be solved. Therefore, the challenge of sanctions might be lower, even turning disadvantages into advantages. The question is whether the food embargo on Russia has only served its purpose. This protectionist policy by Russia was intended to support the domestic agricultural sector, from small businesses to large agricultural companies. With a government initiative to encourage the private sector to invest in the agribusiness industry, the local supply of foodstuff products went through exponential growth since the implementation of the sanctions and started to be exported after a couple of years, so much so that exports have surpassed imports since 2017. 

Our interest focuses on assessing the agricultural sector and examining how sanctions have changed the agricultural business in Russia since 2014. We assess the economic impact of the Russian food embargo on agricultural goods, analysing whether it leads to a window of opportunity for entrepreneurs and investors to take advantage of the business climate, use government subsidies and benefit from different easing policies for domestic production. Our estimations are based on annual data from 2010 to 2020, as later data are not publicly available. The events of 2022 could not help but affect the agricultural sector, although no direct sanctions have been imposed on the industry, even as Russia is subject to numerous sanctions, the highest in its history. The European Commission has repeatedly pointed out that export and import restrictions should exclude products related to medicine, pharmaceuticals, food and agriculture not to harm the population. This may be why only a few bans are aimed directly at the agro-industrial complex.

Nevertheless, sanctions on other sectors have severely affected the agribusiness sector. The economy is a single system, each element involving the others; therefore, this does not mean that the industry is not adversely affected without direct sanctions. At the same time, focused work on the recovery and development of agriculture, the sector was in a relatively strong position by the time the crisis worsened in 2022. In particular, the pig and poultry herds have significantly increased, moving towards self-sufficiency in these types of meat. The gross yield of the main crops in crop production has increased, as demonstrated in the following chapters. 

Section-\ref{sec:two} provides detailed changes in the agricultural industry in Russia. Section-\ref{sec:three}  presents various difference-in-difference estimates based on the role of sanctions on Russian exports and imports, and Section-\ref{sec:four} concludes. Supplementary material on the institutional background and additional results discussed throughout the paper are available in the Appendix.

\section{Russian Agricultural Industry}
\label{sec:two}
\subsection{The state of the agricultural sector in Russia}

As of 2014, Russia's agricultural sector was beginning to emerge from the terrible crisis of the late 1990s. After the collapse of the Soviet Union and the beginning of privatisation, enterprises deprived of the usual state support had to master a completely different, market-based operating environment. Production had fallen sharply over 8 years (livestock production has fallen by almost 60\% and cereals by 52\%) \citep{Rosstat2021}. By the end of the century, most agricultural enterprises were unprofitable. After such a significant decline in the sector, agriculture had a chance after the 1998 default: imported food became more expensive and domestic products became competitive.

Since then, a new phase in developing the country's agricultural sector has begun, Russia has become actively involved in the process of globalization and international integration. There has been an increase in both exports and imports. However, the increase in imports had been many times greater than exports, making Russia dependent on other countries. Therefore, for the first years of the century, Russia was the largest net importer of food. This situation can be described as import dependency: various estimates put imports of food and agricultural products in Russia at between 30\% and 50\% of consumption \citep{Kosenko2015}. This is despite the fact that 20-25\% is already a critical figure, posing a severe economic threat to the country.

Russia has a vast resource potential, which is not sufficiently exploited or not used at all \citep{Deppermann2018, Pylypiv2014}. Not all land is used for its intended purpose, and there is a large amount of ``abandoned land". There are several reasons for this, and perhaps the main reason is the highly skewed structure of farms. In Russia, agroholdings and agrofirms/agrocomplexes have become widespread, combining different-sized agricultural organisations or their branches that have been formed on the site of former agricultural organisations and are under the control of the parent organisation. In some cases, a single organisation with branches is formed. The largest of them control hundreds of thousands of hectares of agricultural land and up to 20,000 agricultural workers. Distortions in land allocation and uneven distribution of agricultural support, strongly focusing on large farms and agriholdings, hinder the development of small farms and prevent them from participating in food value chains, negatively impacting rural development. 

Although household plots produce a third of the country's agriculture and food, the state is shifting its support to them \citep{Uzun2019}. It is important to emphasize that this is a significant but not the only problem. A major challenge is the difficult demographic situation, manifested by an ageing rural population and a lack of human resources in general. According to the \citet{Ozerova2013},  the replacement rate of older people by younger people in rural areas was 238\% in 2000 and is estimated to be 15\% in 2020. There needs to be an effective system for training and retraining professional managerial and production staff for agricultural production. The need for qualified staff prevents many agricultural enterprises from utilizing all their production capacities.

Another problem of Russian agricultural production is its low efficiency. The leading indicator of production efficiency in the economy is labour productivity. In Russian agriculture, labour productivity is several times lower than in leading European countries. For example, in the Netherlands, which is comparable in size to the Moscow region, agricultural productivity is 9 times higher than in Russia, and exports of agricultural products amount to \$90 billion, which is 3 times higher than the results achieved in Russia \citep{Analytical2021}. It is also no secret that many regions lack facilities for processing and storing agricultural products, that they are located at insufficient distances from each other and do not always meet modern requirements. The Russian agricultural sector suffers from outdated technologies, resource efficiency and rational fertilizer use and with conservative consumption behavior \citep{Gokhberg2017}. The country's own science and technology system requires support for both technology development and technology transfer. Add to this the unsatisfactory standard of living of the population engaged in agriculture, and unsurprisingly progress is severely stunted. However, changes are happening and must continue.

\begin{figure}[htbp]
		\caption{Total trade inflows and outflows of agricultural products in Russia (2010-2020)\label{fig:1}}
		\includegraphics[width=.48\textwidth]{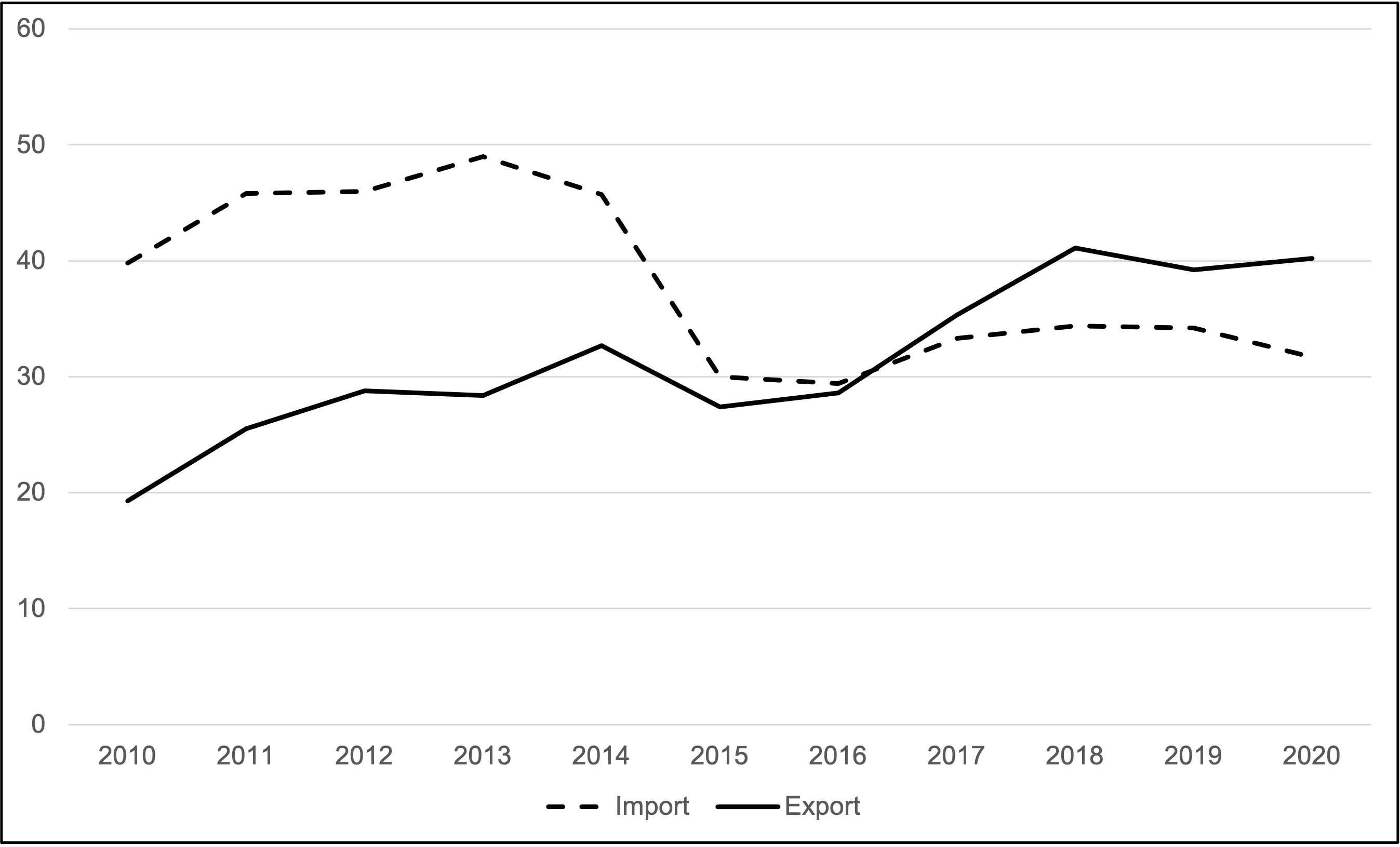}
		\begin{tabular*}{0.45\textwidth}{@{\hskip\tabcolsep\extracolsep\fill}ccccc}
		\multicolumn{4}{p{0.45\textwidth}}{\scriptsize \emph{Notes:} Export and import are measured as a billion USD, data retrieved from the Atlas of Economic Complexity by Harvard’s Growth Lab.}\\ 
		\end{tabular*}
\end{figure}

The implementation of technology is part of the government food security policy. Since agriculture, due to its specific characteristics, is not an independent, self-sufficient industry and cannot develop on the principles of self-sufficiency and self-financing, it requires state support. All of the above circumstances highlight the need to define the role of the state and the extent of its involvement in the management and regulation of agriculture, as well as to develop and implement a set of state support measures. Government support is a prerequisite for the existence and growth of agricultural production and the food market, as the experience of the world's leading powers demonstrates. 

The Russian Food Security Doctrine was introduced in 2010. The programme envisages a comprehensive and accelerated development of the agro-industrial complex and the social infrastructure in rural areas. Only comprehensive support from the state can bring the agro-industrial complex of Russia to a new, highly competitive level \citep{Pylypiv2014}. Since 2010, the policy has become more protectionist, and a path has been taken towards import substitution. The Russian Federation is currently implementing state support measures for producers in the agro-industrial sector aimed at increasing the efficiency and competitiveness of Russian food products. By 2020, support is provided at all stages of the life cycle, from the creation of new products to their promotion, including subsidised lending, subsidised leasing, compensation for transportation and certification costs, and for creation and modernisation of agribusiness facilities, registration of intellectual property abroad, participation in foreign exhibitions and fairs and various business missions. The above-mentioned government interventions have mitigated the possible negative effects of the sanctions and, moreover, have tangibly reduced imports and triggered an increase in exports since 2015 (Figure-\ref{fig:1}).
	
Such radical significant shifts are only possible with government funding, which is increasing relentlessly for the agribusiness sector. Due to the introduction of Russian sanctions, the amount of subsidies to agriculture in 2014 increased 1.5 times, from 165.7 to 252.7 billion roubles, and in total, to 1.8 trillion roubles in 2020 \citep{Semenova2015}. Thus, sanctions and counter-sanctions have triggered increased investment in the agricultural sector.  Over the past 10 years, there has been a significant change in the production, imports and exports of agricultural products as a whole. Whereas these changes were not so evident before 2014, 2014 brought radical transformations, such that, since 2017, imports have surpassed imports for the first time in the ongoing century and have continued to dominate ever since. 

\subsection{Changes in the agricultural industry}
As Russia’s food security model combines state assistance to domestic production while restricting market access, the 2014 embargo became part of this policy \citep{Wegren2016}. The embargo has targeted basic foodstuffs, whose imports have been colossal, and it is these that have been the focus of import substitution. The agricultural products, meat and edible by-products of cattle, pigs and poultry, fish and crustaceans, milk and dairy products, vegetables, fruits and nuts, sausages and similar products made from meat, meat offal or blood; prepared foods made from them, food or finished products manufactured by cheese-making techniques and containing 1.5\% or more of milk fat, salt (from 2016) (including table salt and denatured salt), are prohibited for import into the Russian Federation since 2014. The exceptions are specialised food products for sports nutrition and goods intended for baby food, pharmaceuticals, medicines, medical devices and dietary supplements.

\subsubsection{Meat}
Meat production has undergone a clear positive change in all major breeding areas, namely beef, pork and poultry. Comprehensive work has been carried out to improve production, to open up new countries to Russian livestock products and expand the range of supplies. In 2020, the right of access to 25 foreign countries was obtained for 38 products (in comparison, the right of access to 17 countries for 30 products was obtained in 2019). These include meat and dairy products, as well as feed, non-food products and live animals. In 2020, deliveries of raw meat and finished meat products abroad reached 534,000 tonnes, with 379,000 tonnes exported in the same period last year. There has been a significant decline in imports of raw meat and finished meat products. Imports fell by 22\% in the first 11.5 months of 2020 compared to the same period in 2019 \citep{Analytical2021}.

Poultry production is one of the most stable in the sector. Since 2005, there has been a steady increase in the population. In the year 2017, there was a peak followed by a slight downturn. Even though the poultry population was 518.7 million or 95.2\% of the previous year, compared to 2019, poultry production for slaughter (live weight) on farms of all categories amounted to 6.73 million tonnes, up 0.3\% (+22.4 thousand tonnes) from 2019. It can be assumed that this was due to an increase in the weight of the birds themselves.

After sanctions and counter-sanctions were imposed, poultry imports fell by more than 40 per cent in next year, and have increased in the following years (Figure-\ref{fig:a1}). A significant increase in exports also started only after the introduction of sanctions. Exports have developed considerably, so in 2020, poultry exports ranked first with a share of 35.6\% (295,900 tonnes or USD 429.5 million) in the structure of animal product exports. Poultry meat was exported to 39 countries in 2020, and export volumes increased from 211,000 to 276,000 tonnes compared with 2019 \citep{Analytical2021}. These results have been achieved not only by ramping up production but also by expanding distribution markets. For example, in 2020,  China expanded the list of supplied poultry by-products (turkey by-products) and the list of enterprises eligible to supply poultry meat and by-products to 49 enterprises \citep{Analytical2021}. All this led to exports exceeding imports for the first time by as much as 43 per cent. 

The cattle population, on the other hand, has not been as stable over the period under review, in fact, there has been a sharp decline in heads since 2011 until 2016. In the next 4 years, the population was consistently low. The reduction in the number of livestock in agricultural organisations was largely due to the intensification of production and the renewal of the herd with more productive livestock. The general tendency towards urbanisation and the difficult economic situation of the households also had an impact. Nevertheless, from 2019 the population declines again and continues up to and including 2022. The fall in the 2021 population was the highest since 2014 \citep{Rosstat2021}.

Frozen beef has dominated the Russian trade balance over fresh beef over the past decades. Frozen beef imports before 2014 were 4 to 8 times higher than imports of fresh beef, depending on the year (Figure-\ref{fig:a2}). Since 2014, due to a rapid decline in frozen beef imports to almost fresh beef levels, this ratio has fallen to an average of 2. Thus, in 6 years, imports in both categories decreased twofold and threefold for fresh and frozen beef, respectively.

Beef exports have developed very dynamically since 2010. Since 2016 (two years after the embargo), both groups have seen steady growth, but not comparable to the change from 2019 to 2020; exports have tripled. Overall, exports of fresh and frozen beef have increased by 137 and 120 per cent, respectively, over the period under consideration, and 12 and 8 times since the sanctions were imposed. Chilled beef exports have increased by a third since 2016. In 2020, total beef exports increased from 5,900 tonnes to 14,500 tonnes \citep{Analytical2021}. Beef exports thus occupied the third position in the structure of meat exports with a share of 7.3\%. Such figures directly support the hypothesis that the country has been able not only to meet the needs of its citizens but also to expand the boundaries of consumption to new countries.

In 2020, Russian beef was exported to 19 countries due to the opening up of new countries to Russian products. For example, China has received access to beef (chilled and frozen), and an agreement has been reached to expand the list of beef by-products allowed for export (2 enterprises have been certified). As for Vietnam, there are 4 certified Russian meat processors for beef. Two of them are mixed: 1 for poultry/beef and 1 for pork/beef \citep{Analytical2021}.

As for pork, it is the most successful example of import substitution and the impact of the embargo on the country's food security. Pork imports were increasing in the Russian Federation until 2012, but since 2012 there has been a sharp drop in imports. Up to 2019, imports have only declined, except for a slight increase in 2017. The reduction was by a record 2,1 billion USD or by 252 times. Thus pork imports have fallen almost to zero (Figure-\ref{fig:a3}).

In 2020, pork exports were Russia's second-largest total exports of meat and dairy products. Pork accounted for 27.9\% with exports of 200,300 tonnes. Russian pork was exported to 15 countries in 2020 \citep{Analytical2021}. Since 2014, exports have increased 32-times and exceeded imports for the first time in history. It was exceeded by as much as 28 times, a clear indication that dependency is decreasing.

\subsubsection{Wheat and cereals}
Regarding wheat and cereals, these are the only commodity not subject to embargo. However, it is only possible to mention them as they are in the leading position of agricultural exports at the moment and even in the top 5 of all exports from Russia after mineral fuels, oil, ferrous metals, pearls and precious metals. \citet{Svanidze2019} shows that physical grain trade flows plays a significant role in Russia; however, the country needs more considerable development in market information and developed commodity futures markets. It is the opposite in the US, where information on commodity futures markets recreates a substantial role in the agricultural sector.

At the end of the century in 1998, cereal production was minimal and unable to meet the population's needs. Production went from rising to falling in the first decade of the 21st century. This continued until production collapsed again in 2010, almost to the level of 1998. In the same year, a doctrine on food security was signed, setting the production level necessary for the country to be self-sufficient. In 2020, production reached 133,000 tonnes, providing the country with 167.6\% of its grain reserves. In 2021, however, production was down again by more than 10\% compared to the previous year.

Wheat imports show a strong surge in imports in 2013. The country imported a record amount of wheat in a decade (Figure-\ref{fig:a3}). This is probably due to the noticeably low productivity and harvest in 2012, which did not allow the country to provide the necessary volume of wheat the following year. However, the following year, 2014, imports more than halved and continued to decline until the end of the decade. The decline has been low, with imports going up and down, but the reduction has been more significant than the increase, resulting in pleasantly low totals.

Increasing exports have gone hand in hand with an increase in production. Exports fell sharply in 2013, rose in 2014 and fell again the following year, despite no wheat embargo. Exports have only increased since 2015 and peaked in 2018. After a decline in 2019, exports rose again in 2020, amounting to 49,893 thousand tonnes, 38,554 of which were wheat (in 2019, these figures were 40,670 and 31,896 respectively) \citep{Analytical2021}. 

Therefore the country has become ``the world's top wheat exporter, supplying 20–23 percent of total world exports in 2017-2018" \citep{liefert2020}. The market continues to grow, and Russian grain was shipped to 138 countries in 2020. 

\begin{table*}[htbp]
\def\sym#1{\ifmmode^{#1}\else\(^{#1}\)\fi}
\caption{Differences-in-differences model of trade flows into and from Russia}
\begin{center}
{\footnotesize
\begin{tabular}{p{25mm} c c c c c c}
\hline
            &\multicolumn{1}{c}{Import}&\multicolumn{1}{c}{Import}&\multicolumn{1}{c}{Export}&\multicolumn{1}{c}{Export}&\multicolumn{1}{c}{Net export}&\multicolumn{1}{c}{Net export}\\
            &\multicolumn{1}{c}{(1)}&\multicolumn{1}{c}{(2)}&\multicolumn{1}{c}{(3)}&\multicolumn{1}{c}{(4)}&\multicolumn{1}{c}{(5)}&\multicolumn{1}{c}{(6)}\\
\hline
Sanctions$\times$1 (t=2014) &     0.002             &                     &       0.405\sym{***}&                     &       0.357\sym{***}&                     \\
                            &    (0.074)           &                     &     (0.120)         &                     &     (0.130)         &                     \\
Sanctions$\times$1 (t=2015) &      -0.510\sym{***}  &                     &       0.475\sym{***}&                     &       0.940\sym{***}&                     \\
                            &    (0.092)           &                     &     (0.155)         &                     &     (0.160)         &                     \\
Sanctions$\times$1 (t=2016) &      -0.571\sym{***}  &                     &       0.654\sym{***}&                     &       1.179\sym{***}&                     \\
                            &     (0.101)           &                     &     (0.190)         &                     &     (0.219)         &                     \\
Sanctions$\times$1 (t=2017) &      -0.354\sym{***}  &                     &       0.826\sym{***}&                     &       1.117\sym{***}&                     \\
                            &    (0.099)           &                     &     (0.196)         &                     &     (0.226)         &                     \\
Sanctions$\times$1 (t=2018) &      -0.394\sym{***}  &                     &       0.867\sym{***}&                     &       1.171\sym{***}&                     \\
                            &     (0.123)           &                     &     (0.198)         &                     &     (0.246)         &                     \\
Sanctions$\times$1 (t=2019) &      -0.340\sym{***}  &                     &       1.015\sym{***}&                     &       1.301\sym{***}&                     \\
                            &     (0.128)           &                     &     (0.205)         &                     &     (0.267)         &                     \\
Sanctions$\times$1 (t=2020) &      -0.405\sym{**}   &                     &       1.313\sym{***}&                     &       1.673\sym{***}&                     \\
                            &     (0.197)           &                     &     (0.216)         &                     &     (0.346)         &                     \\
Sanctions                   &                       &      -0.368\sym{***}&                     &       0.791\sym{***}&                     &       1.376\sym{***}\\
                            &                       &     (0.102)         &                     &     (0.169)         &                     &     (0.289)         \\
\hline
Constant                    &       16.67\sym{***}  &       16.67\sym{***}&       15.33\sym{***}&       15.33\sym{***}&      -1.407\sym{***}&      -1.096\sym{***}\\
                            &     (0.161)           &     (0.161)         &     (0.175)         &     (0.175)         &     (0.172)         &     (0.169)         \\
\hline
Observations                &        3097           &        3097         &        3052         &        3052         &        3041         &        3041         \\
Clusters                    &        286            &        286          &        284          &        284          &        281          &        281          \\
R2                          &      0.0997           &      0.0995         &      0.0545         &      0.0399         &      0.0696         &      0.027         \\
\hline
\end{tabular}}
\vspace{1mm}
\begin{tabular*}{0.65\textwidth}{@{\hskip\tabcolsep\extracolsep\fill}cccccc}
	\multicolumn{4}{p{0.62\textwidth}}{\scriptsize \emph{Notes:}  The first panel shows the effect of sanctions in a particular year (i.e., the interaction of sanctions with the year dummies). Columns (1), (3) and (5) refer to models in which we consider different sanction effects in each year, while columns (2), (4) and (6) correspond to the case when we assume the same sanction effect across all years. In all models we control for product, as well as time fixed effects. Standard errors clustered at the exporter level are reported in parentheses. Significance codes: * 10\%, ** 5\%, *** 1\%.}\\
 \end{tabular*}
 \end{center}
\end{table*}

The main countries importing Russian grain are: \\
Turkey - 18.6\% (9048.6 thsnd tonnes, or \$1,913 mln); Egypt - 17.5\% (8258.9 thsnd tonnes, or \$1,797.6 mln); Saudi Arabia - 5.5\% (3058.4 thsnd tonnes, or \$568.3 mln); Bangladesh - 4\% (1943.8 thsnd tonnes or \$409.5 mln); Azerbaijan - 3\% (1,525 thsnd tonnes or \$312.9 mln). \citep{Analytical2021}

The structure of grain exports in 2020 is as follows: wheat - 80.7\% (38,878.2 thsnd tonnes, or \$8284.8 mln); barley - 10.8\% (6,118.8 thsnd tonnes, or \$1108 mln); maize - 7.2\% (4,021.1 thsnd tonnes, or \$741.5 mln); rice - 0.7\% (148.3 thsnd tonnes, or \$69 mln); buckwheat, millet, other - 0.4\% (105.3 thsnd tonnes, or \$41.5 mln).

\section{Data and Methodology}
\label{sec:three}

\subsection{Data}

To verify whether the sanctions had a positive impact on the export and import transactions of the Russian Federation, we use data from the Atlas of Economic Complexity by Harvard Growth Lab from 2010 to 2020. This system was chosen because of its extreme convenience and completeness of data, as the data classification corresponds to an international classification system, namely the Harmonized System (HS, 1992). This dataset allows us to analyse around 280 product groups from the agricultural category for each year. The number of export and import groups varied each year slightly as there needed to be data provided. Nevertheless, the overall total varied around 280 (279 to 285 on average). In this way, we separated the sanctions groups from the control groups, allowing us to show the changes taking place fully.

\subsection{Methodology}

We analyze agricultural trade flow data using a difference-in-difference model. The model is specified as a panel regression with a fixed time effect, and clustered by product groups. Our baseline model follows \citet{Belin2021}:
\begin{equation}
    Y_{it} = \sum^{2017}_{s=2011} \beta_s Sanctions \times 1(t=s) + \alpha_i + \alpha_t + \epsilon_{ij}
\end{equation}
where $Y_{it}$ are the imports of agricultural product $i$ in year $t$. Additionally, we estimate:
\begin{equation}
    Y_{it} = \beta^{} Sanctions \times 1(t \geq 2014) + \alpha^{*}_i + \alpha^{*}_t + \epsilon^{*}_{ij}
\end{equation}
where time fixed effects are interacted with a Sanctions dummy taking a value of one if a given imported product is subject to a sanctioning regime and zero otherwise. $\beta_s$ represents the yearly average differences between the sanctioned and control group.

\section{Results}\label{sec:four}

Russian agricultural trade flows that are subject to the Russian counter-sanction were modelled by difference-in-difference specifications (1) and by its simplified version (2). We assess the same specifications for Russian export and net export as well. All six results are reported in Table-1. We did not present the interaction coefficient due to space savings. The model demonstrates a significant reduction in imports of sanctioned products compared to the control variables. In parallel, there has been a considerable increase in exports, including net exports.  

The main reason contributing to these changes is the introduction of counter-sanctions, namely the food embargo. The imposition of European sanctions on Russia has only served to accelerate action to reduce dependence on Western supplies. After all, dependency is a lever of pressure and a way of control that Russia has been trying to get rid of for decades. Dependence on imports is seen as a danger and makes Russia vulnerable \citep{Korhonen2018}. The mere fact that Russia has directed its sanctions at food shows a desire to reduce food dependency \citep{Wengle2016}.

While the European sanctions were taken from a Western perspective, they completely ignored the country's national interest. Thus sanctions on Russia either had little or no significant impacts \citep{Wegren2016}. \citet{Belin2021} emphasize that Western sanctions were less restrictive than Russian counter-sanctions, allowing Western exporters to ship sanctioned product due to prior contracts before sanction imposition. On the contrary, Russian sanctions were more restrictive due to strong customs duty incentives. Meanwhile, Western exporters were using third countries to avoid sanction imposition, which is extremely difficult to track and analyse. It is this restrictive and rigid policy that has led to such impressive results.

\begin{figure}[htbp]
	\begin{center}
		\caption{Differences-in-Differences plots of trade inflows into Russia \label{fig:2}}
		\includegraphics[width=.48\textwidth]{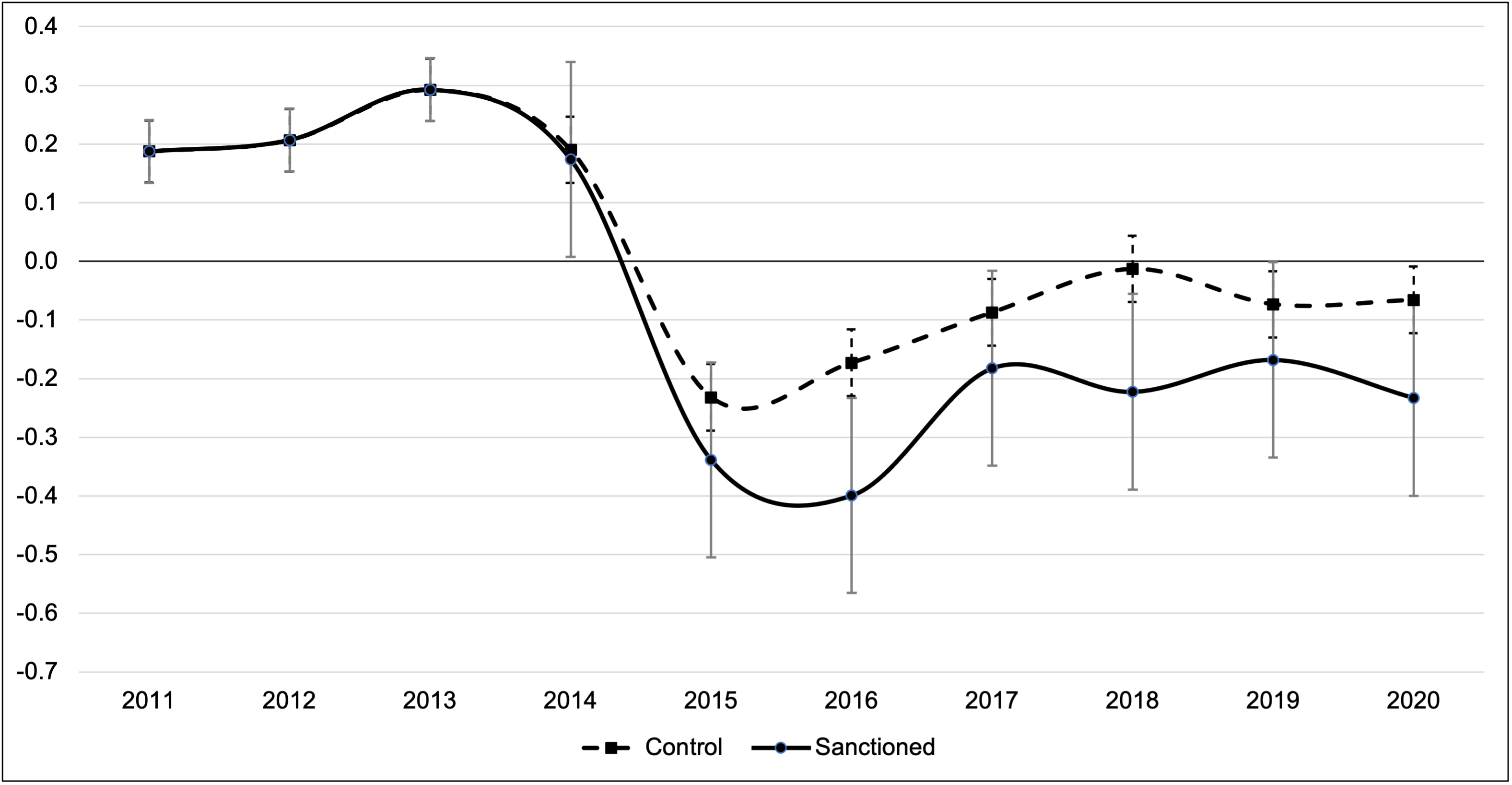}
		\begin{tabular*}{0.45\textwidth}{@{\hskip\tabcolsep\extracolsep\fill}ccccc}
			\multicolumn{4}{p{0.45\textwidth}} {\scriptsize \emph{Notes:} Differences-in-Differences plots indicating the mean trade inflows into Russia with 95\% confidence intervals.The first year of sanctions is 2014.}\\ 
		\end{tabular*}
	\end{center}
	\vspace{-0.4cm}
\end{figure}

Figure-\ref{fig:2} plots results from differences-in-differences estimation of trade inflows into Russia. It shows that the sanctions have had a significant negative impact on foodstuff imports. Imports have almost halved in the first two years since sanctions were imposed. Comparing the sanctions and control groups, there is a clear difference between them, with the fall in the sanctions groups being by far the highest. Subsequent fluctuations were not as critical and import reductions occurred in both groups, although not as dramatic as in the first two years.

Counter-sanctions have obviously contributed to the decline in imports to a large extent, but this is not the only reason. In 2003, for example, the state imposed a restrictive meat import quota regime, which has not yet been abolished. The government has also begun widely imposing sanitary and hygiene restrictions on imports of agricultural products, a policy that has not abated since Russia joined the World Trade Organisation in 2012 \citep{liefert2020}.

A sharp drop in imports and a lack of supply of familiar products has stimulated local producers to act decisively in order to fill the empty niche and the previous need for imports. The absence of foreign competition and subsidised loans have attracted many new players to agricultural markets. This inevitably generates a new interest among some businessmen in the protection and development of domestic agriculture \citep{Korhonen2018}. Not only have agrarians begun to receive comprehensive state support, but they have also become protected by the state's protectionist policies, which impose the aforementioned import quotas and then tariff-rate quotas \citep{Wegren2017}.

\begin{figure}[htbp]
	\begin{center}
		\caption{Differences-in-Differences plots of trade outflows from Russia \label{fig:3}}
		\includegraphics[width=.48\textwidth]{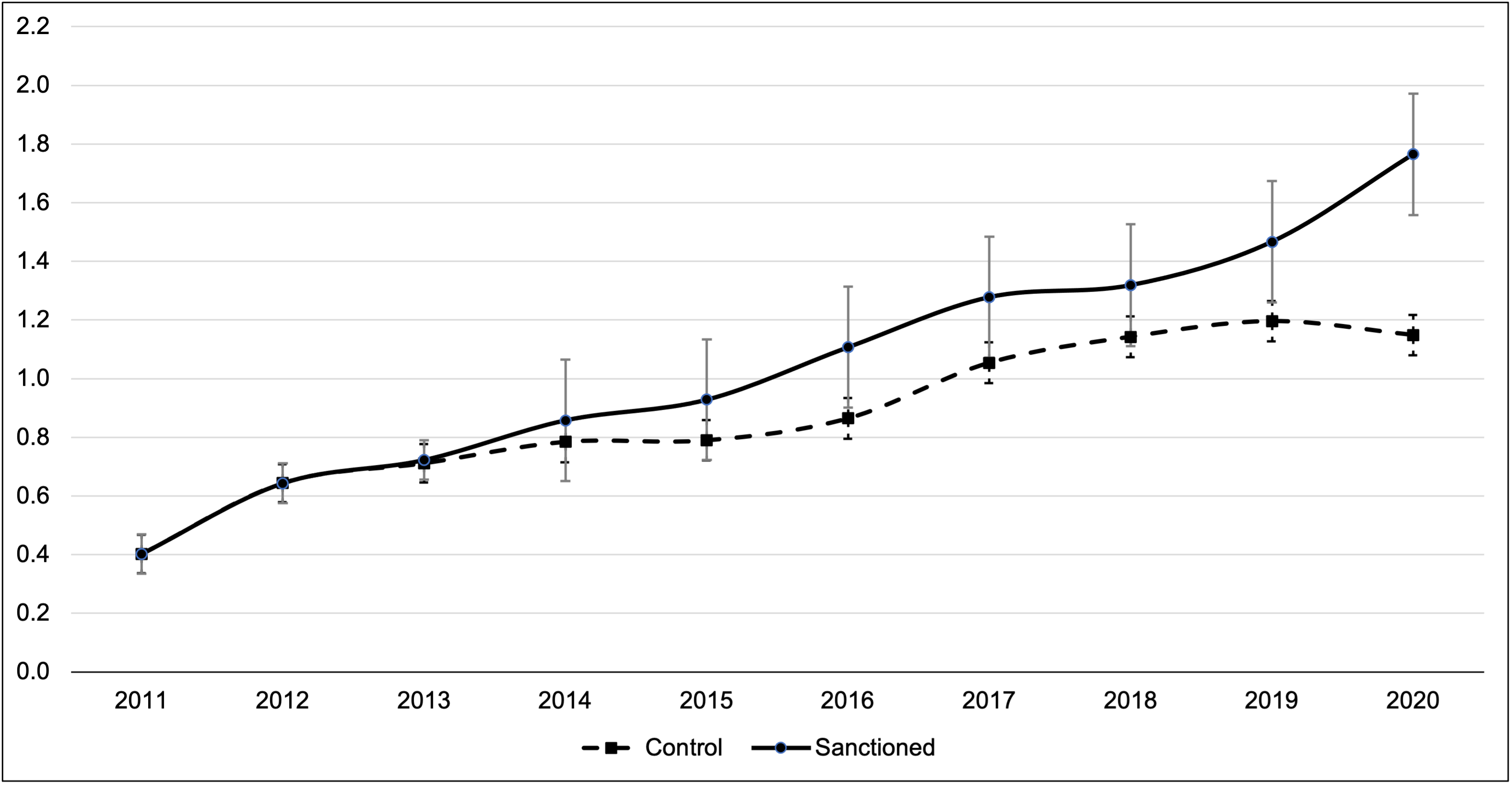}
		\begin{tabular*}{0.45\textwidth}{@{\hskip\tabcolsep\extracolsep\fill}ccccc}
			\multicolumn{4}{p{0.45\textwidth}} {\scriptsize \emph{Notes:} Differences-in-Differences plots indicating the mean trade inflows into Russia with 95\% confidence intervals.The first year of sanctions is 2014.}\\ 
		\end{tabular*}
	\end{center}
	\vspace{-0.4cm}
\end{figure}

Figure-\ref{fig:3} presents results from differences-in-differences estimation of trade outflows from Russia. We observe that food exports have increased over the last decade. Feasible shifts have started taking place in the post-sanctions period, with exports only growing yearly. Indeed, it is challenging to underestimate government support on this issue. Active agricultural assistance is the only key to achieving food and, by extension, national security. With state aid, the Russian food sector can be stronger than before the embargo, meaning that food insecurity in the traditional sense will remain a minor problem \citep{Wegren2017}.

According to an analysis by \citet{Shik2020}, unsurprisingly, until 2019, the main goal was to increase production to the values set out in the food security doctrine. More than 40\% of the funds were allocated to programmes to expand production. And from 2019 onwards, the course is set to boost exports. This was only possible by meeting the needs of the domestic population. According to the national report of the Ministry of Agriculture, the targets set in the new food security doctrine have been achieved for all indicators except milk and potatoes. Budget support has had a significant positive impact on agricultural productivity growth and further export expansion. However, the state still seeks more effective and equitable instruments to support it.

Concerning the development of exports during this time, unexpected heights have been achieved, and the results of the combination of the economic situation, state support and public initiative can already be seen. Since 2017, exports have finally begun to prevail over imports up to and including 2020, and the balance of payments is also eventually positive. Most subsectors have since managed to reorient their exports outside the EU to other markets \citep{Korhonen2018}. By 2020 the geography of Russian agribusiness exports encompasses virtually all countries worldwide. The main export destinations are the CIS (23\%), the Middle East (20\%), the Asia-Pacific region without China (15\%), Africa (14\%), Europe and China (13\% each) \citep{Analytical2021}. Our findings confirm \citet{Korhonen2018} - the Russian food embargo has triggered a decrease in imports of a large number of food products and stimulated a reciprocal increase in export potential.

\section{Conclusion}
\label{sec:five}

The aim of our research was to determine whether and how the introduction of economic sanctions in 2014 affected the development of the agricultural sector in the Russian Federation. Until 2014, Russia was significantly dependent on food imports; therefore, since 2010, Russia has embarked on a path to reduce dependence on food imports. Imports of certain categories of goods varied between 30 \%, and 50 \%, although 20 \% is already considered critical. In this regard, in 2010, the first food security doctrine was introduced, which outlined a set of measures to ensure the country's food sovereignty. The strategic objective of food security is to provide the population of the country with safe agricultural products. It must be guaranteed by the stability of domestic production and the availability of the necessary reserves and stocks.

Russia was hit by a wave of criticism and sanctions over Russia's annexation of Crimea. As a response to Western sanctions, the Russian Federation subsequently imposed an embargo on agricultural products from sanction-supporting countries. Despite the aggravated political situation, the food embargo has been a catalyst for the country's agricultural development, as losses had to be made up as quickly as possible. The loss of imported goods has encouraged local producers to fill the missing market niches. The producers alone could not have met such a huge demand, and government support was needed. Gradually, the state has started to expand its support for agrarians, and by 2020 support is being provided at all stages of production. The state budget, allocated for equipment modernisation, production incentives, reimbursement of certification and transport costs, played a pivot role in mitigating the negative impact of the sanction. The result of joint activities between farmers and the state is the achievement of a satisfactory level of self-sufficiency in essential import-substituting products.

A difference-in-difference model using trade flows of Russian agricultural products from 2010 to 2020 shows a significant reduction in imports of sanctioned products compared to the control variables. In parallel, there is a significant increase in exports, including net exports. Thus, it can be concluded that the Russian embargo has not had a significant impact, indeed has even spurred the development of agriculture in the Russian Federation. Russia has provided itself with basic foodstuffs and reduced its dependence on imports. Moreover, domestic demand has been followed by an increase in export potential. Therefore,  sanctions have not been effective in deterring in the agricultural sector. 

\bibliographystyle{elsarticle-harv} 
\bibliography{Literature}

\newpage

\section*{Appendix}
\label{sec:appendix}

\renewcommand{\thesection}{A\arabic{section}}%
\renewcommand{\thetable}{A\arabic{table}}%
\renewcommand{\thefigure}{A\arabic{figure}}%
\renewcommand{\theequation}{A\arabic{equation}}%
\setcounter{equation}{0}%
\setcounter{table}{0}%
\setcounter{figure}{0}

\begin{figure}[htbp]
	\begin{center}
		\caption{Import and export of poultry in Russia (2010-2020) \label{fig:a1}}
		\includegraphics[width=.45\textwidth]{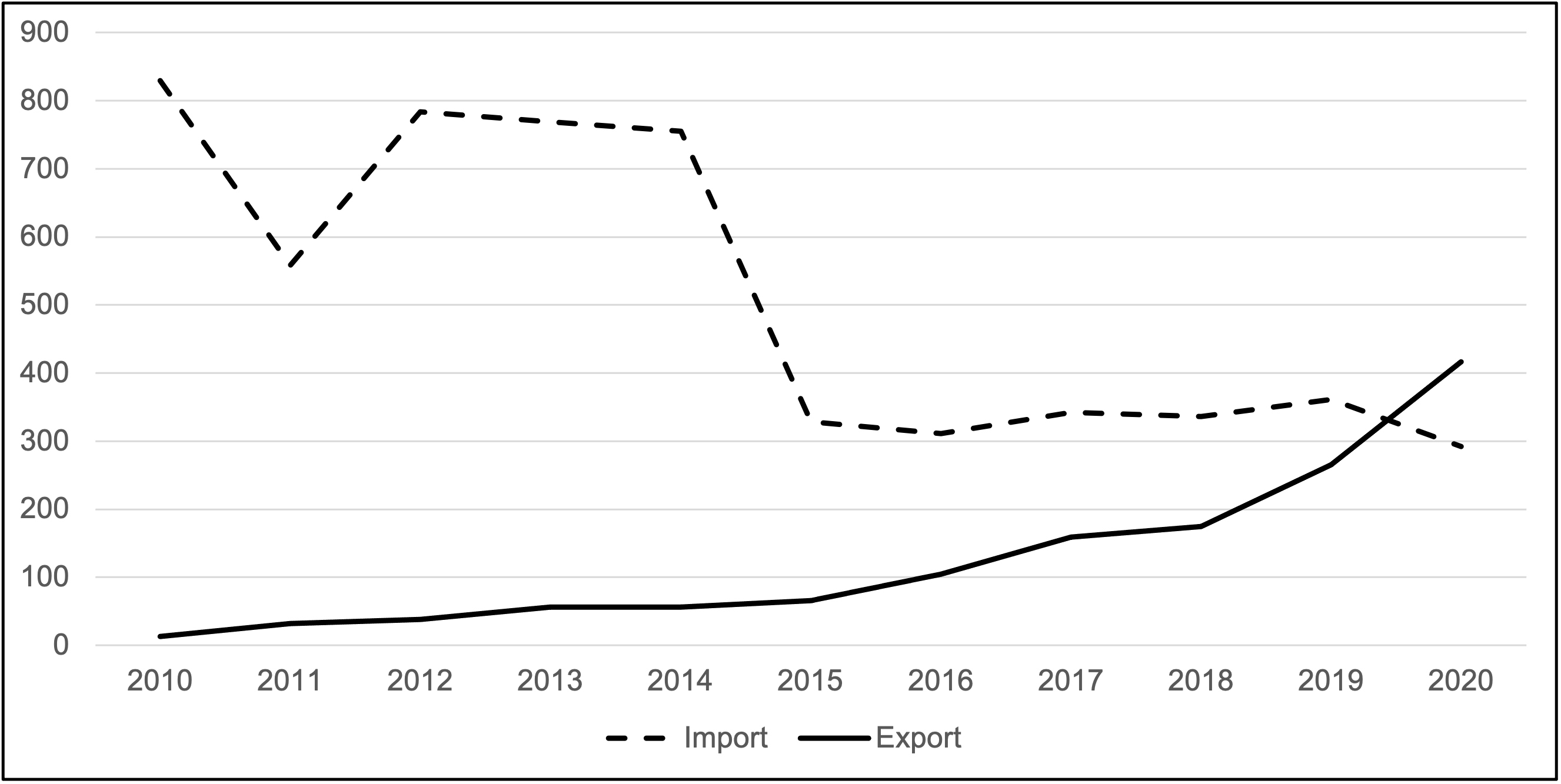}
		\begin{tabular*}{0.45\textwidth}{@{\hskip\tabcolsep\extracolsep\fill}ccccc}
			\multicolumn{4}{p{0.42\textwidth}} {\scriptsize \emph{Notes:} Export and import are measured as a million USD, data retrieved from The Atlas of Economic Complexity by Harvard’s Growth Lab.}\\ 
		\end{tabular*}
	\end{center}
	\vspace{-0.4cm}
\end{figure}

\begin{figure}[htbp]
	\begin{center}
		\caption{Import and export of fresh and frozen beef in Russia (2010-2020) \label{fig:a2}}
		\includegraphics[width=.45\textwidth]{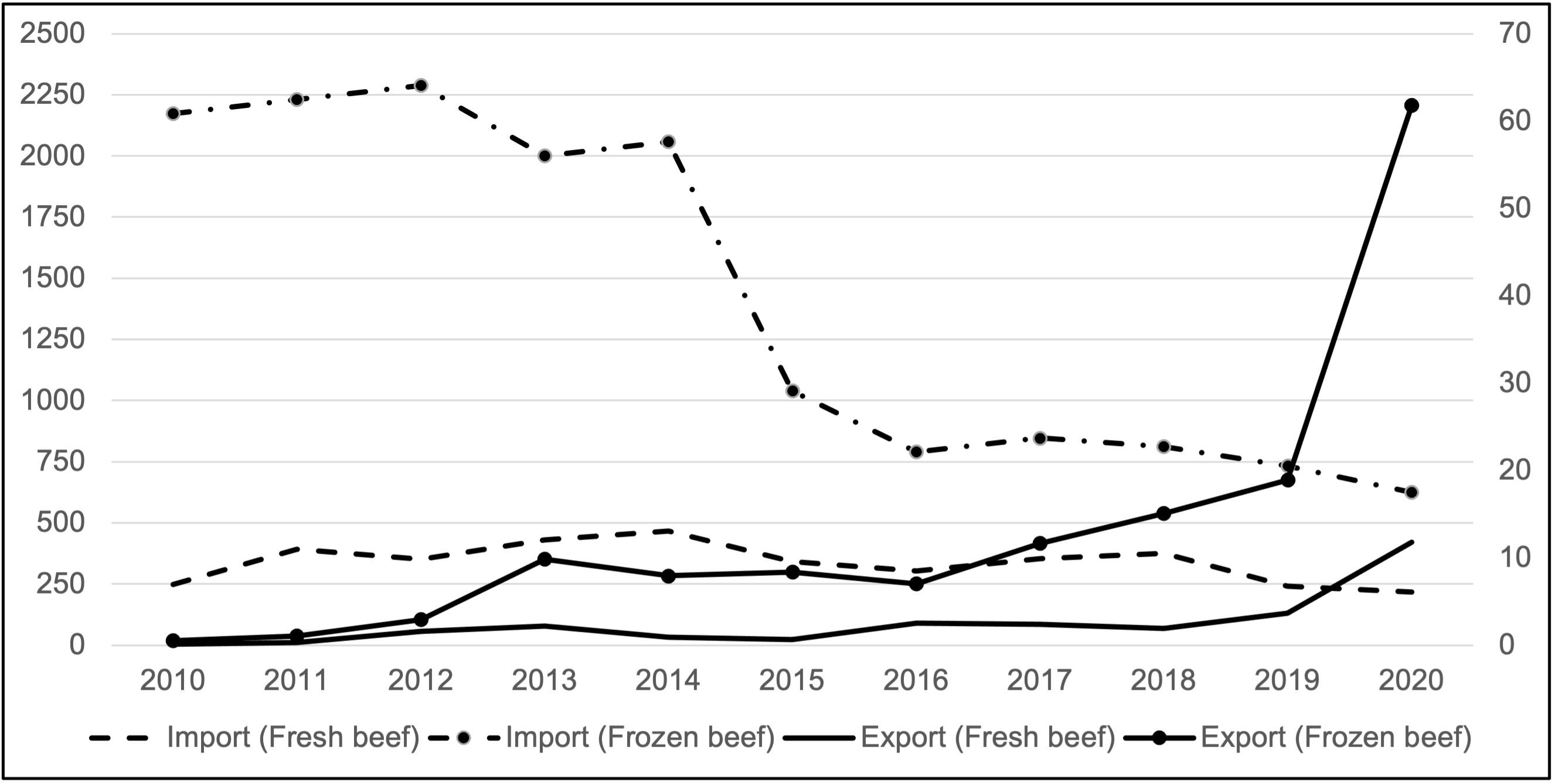}
		\begin{tabular*}{0.45\textwidth}{@{\hskip\tabcolsep\extracolsep\fill}ccccc}
			\multicolumn{4}{p{0.42\textwidth}} {\scriptsize \emph{Notes:} Export and import measured as a million USD, data retrieved from the Atlas of Economic Complexity by Harvard’s Growth Lab. The left axis represents import values, and the right axis represents export values. }\\ 
		\end{tabular*}
	\end{center}
	\vspace{-0.4cm}
\end{figure}

\newpage

\begin{figure}[htbp]
	\begin{center}
		\caption{Import and export of pork in Russia (2010-2020) \label{fig:a3}}
		\includegraphics[width=.45\textwidth]{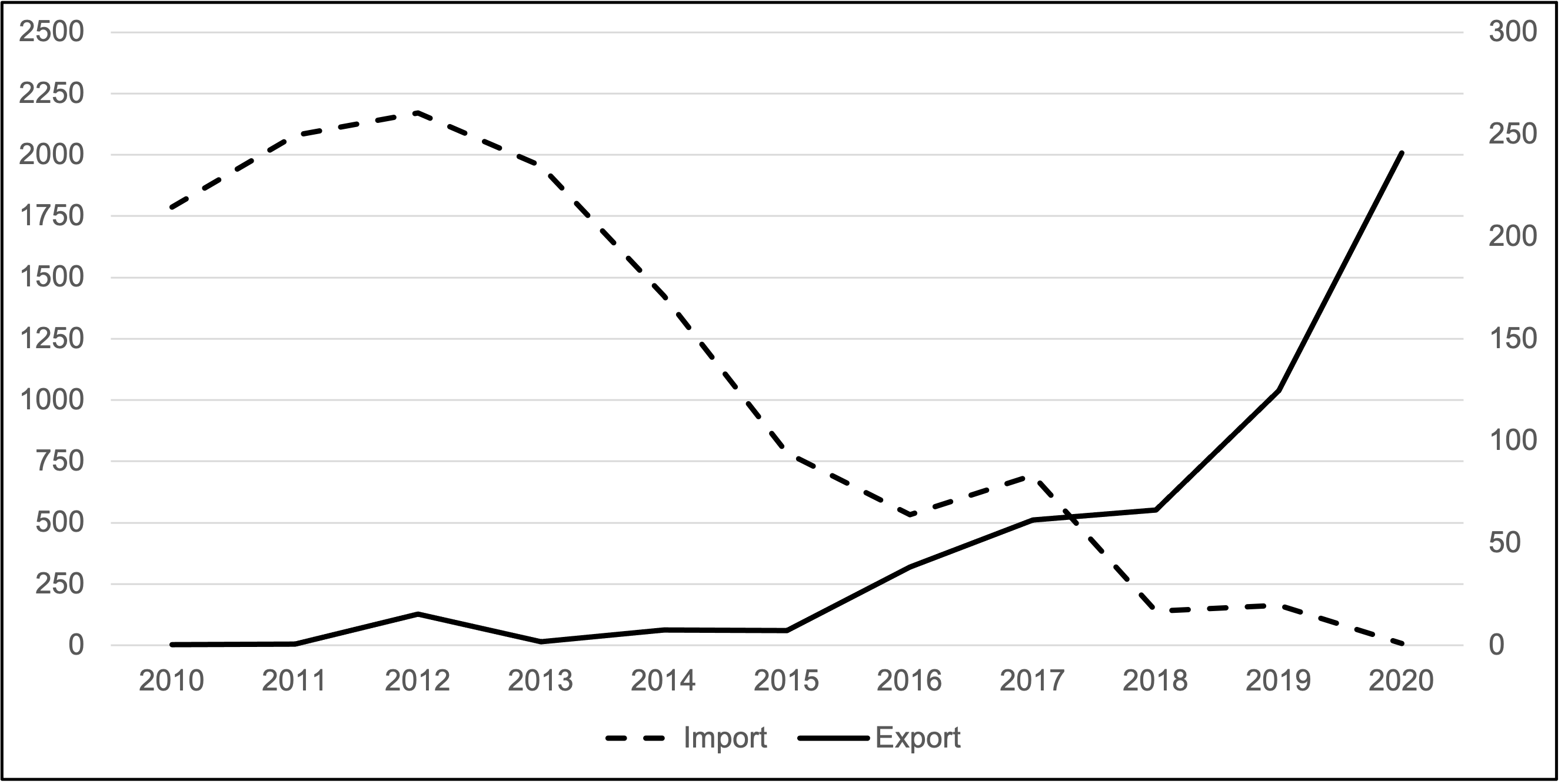}
		\begin{tabular*}{0.45\textwidth}{@{\hskip\tabcolsep\extracolsep\fill}ccccc}
			\multicolumn{4}{p{0.42\textwidth}} {\scriptsize \emph{Notes:} Export and import measured as a million USD, data retrieved from the Atlas of Economic Complexity by Harvard’s Growth Lab. The left axis represents import values, and the right axis represents export values.}\\ 
		\end{tabular*}
	\end{center}
	\vspace{-0.4cm}
\end{figure}

\begin{figure}[ht]
	\begin{center}
		\caption{Import and export of wheat and meslin in Russia (2010-2020) \label{fig:a4}}
		\includegraphics[width=.45\textwidth]{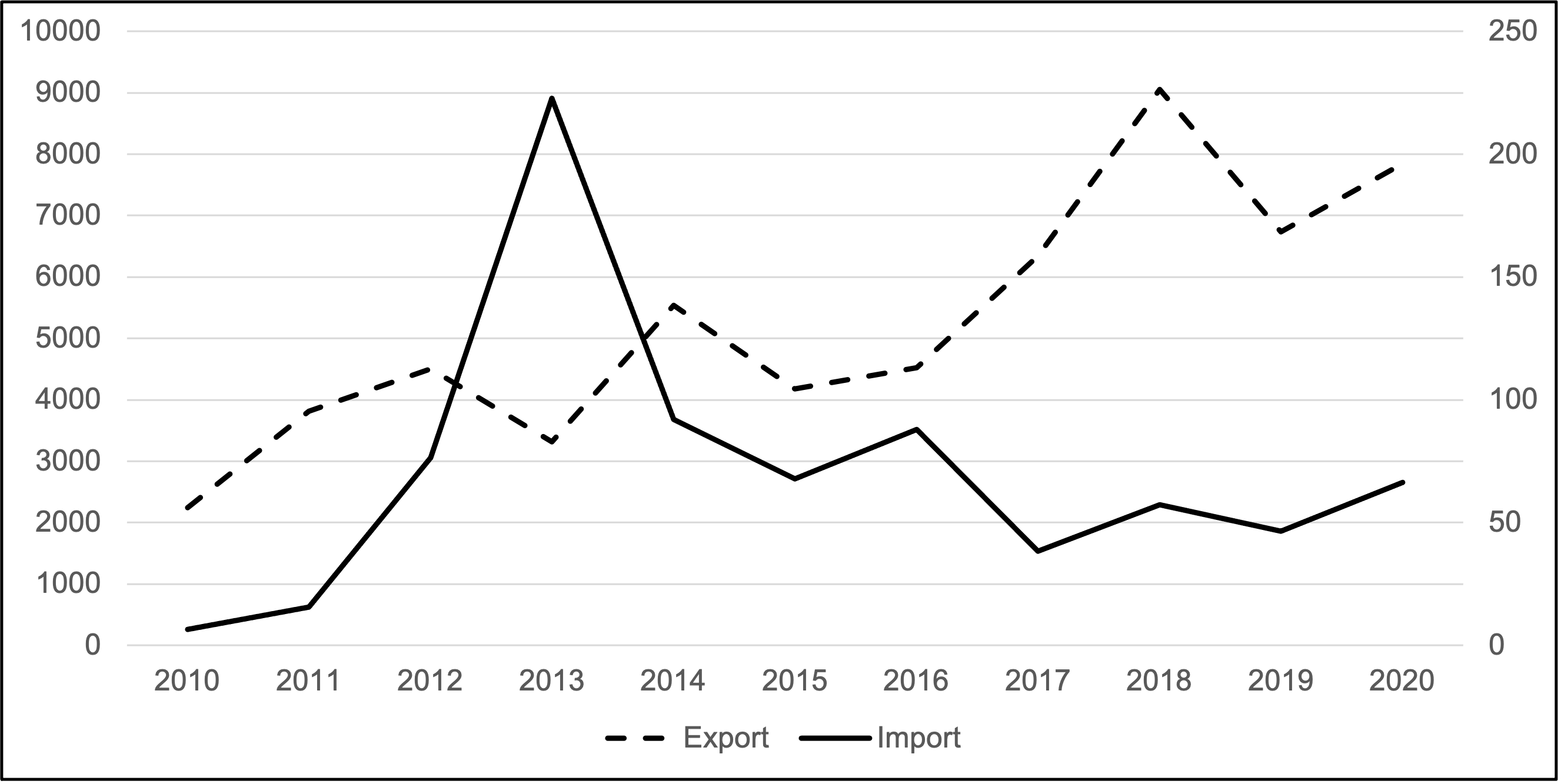}
		\begin{tabular*}{0.45\textwidth}{@{\hskip\tabcolsep\extracolsep\fill}ccccc}
			\multicolumn{4}{p{0.42\textwidth}} {\scriptsize \emph{Notes:} Export and import measured as a million USD, data retrieved from the Atlas of Economic Complexity by Harvard’s Growth Lab. The left axis represents import values, and the right axis represents export values.}\\ 
		\end{tabular*}
	\end{center}
	\vspace{-0.4cm}
\end{figure}

\end{document}